# STEPIC: High-Speed Imaging via Spatio-Temporal Encoding in Photonic Integrated Circuits

ANDREA CICERI,[1] GIACOMO CORRIELLI,[2] GIULIA BERTOLINI,[3] CINZIA DE MARCO,[3] VERA CAPPELLETTI,[3] SERENA DI COSIMO,[3] YUNJIE DENG,[4] YUQI ZHOU,[4] MARTINA RUSSO,[5] HIROSHI KANNO,[4] KELVIN LEE,[4] TIANBEN DING,[4] ANDREA BASSI,[1] ROBERTO OSELLAME,[2] FRANCESCA BRAGHERI,[2] KEISUKE GODA,[4,6,7,8,9,10] NADIA BRANCATI[5] AND PETRA PAIÈ,[1,*]

[1]*Dipartimento di Fisica, Politecnico di Milano, 20133 Milano, Italy*
[2]*Istituto di Fotonica e Nanotecnologie, Consiglio Nazionale delle Ricerche, 20133 Milano, Italy*
[3]*Fondazione IRCCS Istituto Nazionale dei Tumori di Milano, 20133 Milano, Italy*
[4]*Department of Chemistry, University of Tokyo, Tokyo, 113-0033, Japan*
[5]*Istituto di calcolo e reti ad alte prestazioni, Consiglio nazionale delle ricerche, 80131 Napoli, Italy*
[6]*SiRIUS Institute of Medical Research, Tohoku University, Miyagi, 980-0872, Japan*
[7]*Graduate School of Medicine, Tohoku University, Miyagi, 980-0872, Japan*
[8]*International Centre for Synchrotron Radiation Innovation Smart, Tohoku University, Miyagi, 980-8577, Japan*
[9]*Institute of Technological Sciences, Wuhan University, Wuhan, Hubei, 430072, China*
[10]*Department of Bioengineering, University of California, Los Angeles, California, 90095, USA*

*petra.paie@polimi.it

**Abstract:** High-speed imaging of cells in flow is essential for probing cellular heterogeneity in large populations. Existing imaging approaches based on single-pixel detection and spatio-temporal encoding provide exceptional speed, but typically rely on bulky free-space optics, long dispersive elements, and are prone to alignment instabilities. Here, we introduce STEPIC Microscopy, the first fully integrated on-chip system for high-speed imaging via spatio-temporal encoding in photonic integrated circuits. Our platform leverages waveguides, splitters, fiber delay-lines, and 3D optical remappers to encode spatial information into the temporal domain, enabling robust image reconstruction of cells flowing through microchannels. The monolithic architecture provides a compact and robust platform for high-throughput bioimaging, enabling scalable and practical implementations of ultrafast imaging systems.

## 1. Introduction

High-speed imaging techniques are increasingly important for studying dynamic biological processes and for analyzing large populations of microscopic objects. In the latter case, probing cellular heterogeneity requires technologies capable of rapidly interrogating large populations while preserving spatial and morphological information at the single-cell level [1,2]. Flow cytometry has emerged as a widely used tool for analyzing large cell populations as they flow through a detection region. However, traditional flow cytometers rely on fluorescence readouts, providing multiparametric biochemical information but lacking direct spatial and morphological characterization [3].

In recent years, imaging flow cytometry has further advanced the field by enabling the acquisition of images for each analyzed cell, thereby enriching the information content [4,5]. The additional morphological and spatial features accessible through imaging significantly improve classification and diagnostic potential. However, this increased information comes at the expense of throughput: image acquisition typically slows down the analysis, limiting its applicability in contexts where millions of cells must be screened.

To overcome these limitations, several ultrafast imaging strategies based on camera detection have been proposed [6–8]. While these approaches enable higher acquisition speeds, they remain constrained by fundamental limitations of camera architectures, such as serial charge-transfer readout, which restricts practical acquisition speeds to about $10^2$–$10^3$ cells $s^{-1}$, or by trade-offs between frame rate and field of view.

In contrast, single-pixel photodetectors provide higher temporal bandwidth, continuous operation, and superior signal-to-noise ratio, enabling ultrafast measurements beyond the reach of camera-based systems [9–12]. Among these approaches, time-stretch microscopy [13,14] has demonstrated line-scan rates in the MHz regime by temporally stretching broadband femtosecond laser pulses and dispersing their frequency components, so that each spectral slice illuminates the sample at a different position and time. While this method achieves microseconds temporal resolution, it requires expensive instrumentation (broadband femtoseconds laser sources, high-bandwidth oscilloscopes), suffers from high optical losses, and relies on kilometers of dispersive fibers, which prevent efficient operation in the visible range where spatial resolution would be higher. More recently, FACED microscopy has been introduced as an alternative ultrafast imaging approach [15]. FACED exploits a pair of nearly parallel mirrors to generate temporally and spatially shifted replicas of a nanosecond laser pulse, eliminating the need for mechanical scanning. The sequence of pulses illuminates the specimen, and the reflected or transmitted signals are recorded by a fast photodiode. By analyzing the intensity fluctuations associated with each pulse, a two-dimensional image of the sample can be reconstructed. Compared to time-stretch microscopy, FACED reduces the need for expensive ultrafast lasers and high-bandwidth oscilloscopes, significantly lowering system costs. Nonetheless, both approaches still rely on free-space optics and delicate alignment, making the setups bulky, complex, and prone to instabilities.

These instabilities represent a major barrier for real-world applications. Day-to-day variations in image quality reduce measurement reproducibility and can strongly affect automated analysis pipelines, particularly those based on machine learning, which require consistent input data. This limitation becomes critical in applications such as rare-cell detection, for instance circulating tumor cells (CTCs) in liquid biopsy workflows [16].

We argue that both setup complexity and instability can be addressed through a fully integrated approach. By monolithically integrating the key optical components onto a photonic chip, free-space alignment is eliminated, resulting in a compact and mechanically robust system that can be operated by non-specialist users.

In this work, we demonstrate the first on-chip high-throughput imaging microscope, which we term STEPIC Microscopy (Spatio-Temporal Encoding in Photonic Integrated Circuits). The device combines an advanced optofluidic system with integrated optics to encode spatial information into the temporal domain. Single-mode waveguides, integrated splitters, and three-dimensional remappers are fabricated via femtosecond laser micromachining, together with fiber-based delay lines for temporal encoding. The same fabrication technology is employed to realize the microfluidic channel in glass. Cells flowing through the microchannel are illuminated by temporally encoded laser pulses, and the transmitted signal is collected by a single-pixel photodetector. Knowledge of the implemented spatio-temporal mapping allows computational reconstruction of cell images at high throughput.

The monolithic integration of the optical components eliminates the need for free-space alignment, providing a compact and mechanically stable platform compared with conventional single-pixel imaging implementations**.** In the current implementation, STEPIC microscopy reconstructs a cell image in approximately 100 µs, corresponding to throughputs on the order of $10^4$ cells $s^{-1}$. This acquisition time is significantly shorter than that reported in previously demonstrated integrated microscopy platforms [17–20], highlighting the potential of the approach for real-time, large-scale biological analysis and point-of-care applications, while benefiting from the simplicity and ease of operation enabled by the integrated architecture.

## 2. Microscope design, fabrication and assembly

The device layout, schematically shown in Fig. 1, was optimized to implement spatio-temporal encoding within a fully integrated architecture, enabling image reconstruction with a single-pixel detector. The system integrates three main functional blocks: pulse replication through an on-chip splitter network, temporal encoding using fiber-based delay lines, and spatial encoding through a three-dimensional optical remapper. These elements operate sequentially to generate a train of spatially and temporally separated illumination pulses. The transmitted signal is subsequently collected by a GRIN lens and focused into a microfluidic channel. All optical and fluidic components are permanently aligned and bonded together, forming a compact monolithic platform.

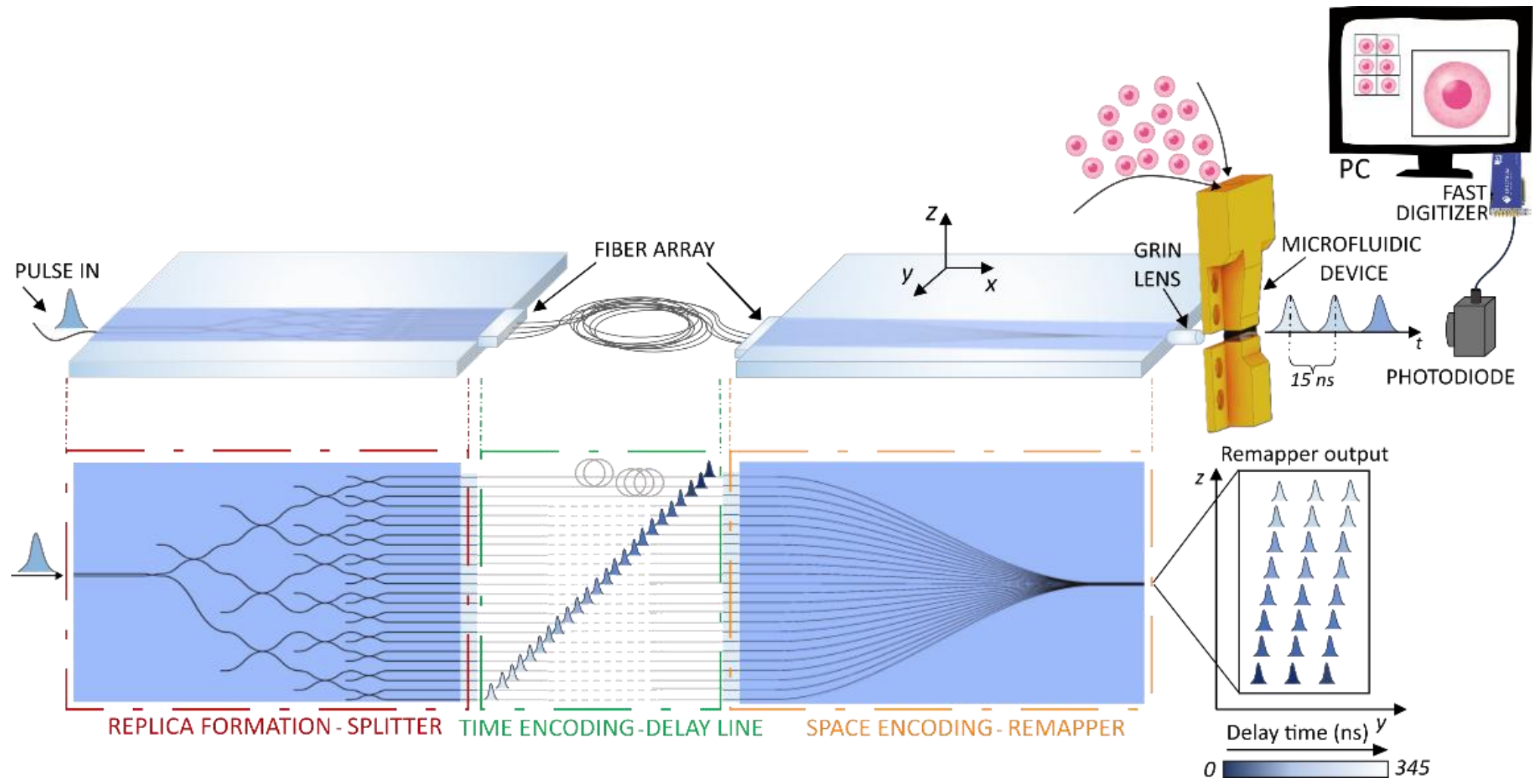

**Fig.1 Layout of the fully integrated STEPIC microscope, combining optical and fluidic elements on a single chip.** A magnified view of the integrated photonic components is shown, comprising: (1) pulse replication through an on-chip beam-splitter network, (2) temporal encoding via fiber-based delay lines, and (3) spatial encoding implemented by a three-dimensional optical remapper. This architecture generates a train of optical pulses that are both temporally and spatially separated. The color coding at the remapper output indicates the relative temporal delay of each spatial mode, ranging from dark blue (short delay) to white (long delay).

The system is driven by a pulsed blue laser (λ = 488 nm) with tunable pulse duration and repetition rate was coupled to the chip through an optical fiber. The fiber was permanently fixed after optimizing the coupling efficiency. In the experiments, the repetition rate was set to 2 MHz, while the pulse duration was set to 10 ns. The chosen repetition rate represents the highest value compatible with the temporal span of the pulse train, avoiding overlap between consecutive bursts, whereas the 10 ns pulse width corresponds to the minimum duration available from the laser source.

The first on-chip element is an integrated splitter network. Using a cascade of evanescent couplers (33:66 the first one, and 50:50 all the others), the input pulse is divided into 24 sub-replicas that will spatially sample the specimen. Once generated, these sub-pulses are manipulated both temporally and spatially by the subsequent elements of the device to achieve the desired spatio-temporal encoding. Temporal encoding is implemented through a fiber-array delay network, in which each fiber introduces an incremental delay of 15 ns, corresponding to a 3 m length difference in glass. This configuration generates a train of 24 temporally separated pulses, repeated at the laser repetition rate of 2 MHz. The 15 ns delay corresponds to an effective intra-burst pulse spacing equivalent to ~66.7 MHz, enabling dense temporal sampling of the specimen while operating with a low repetition-rate laser source.

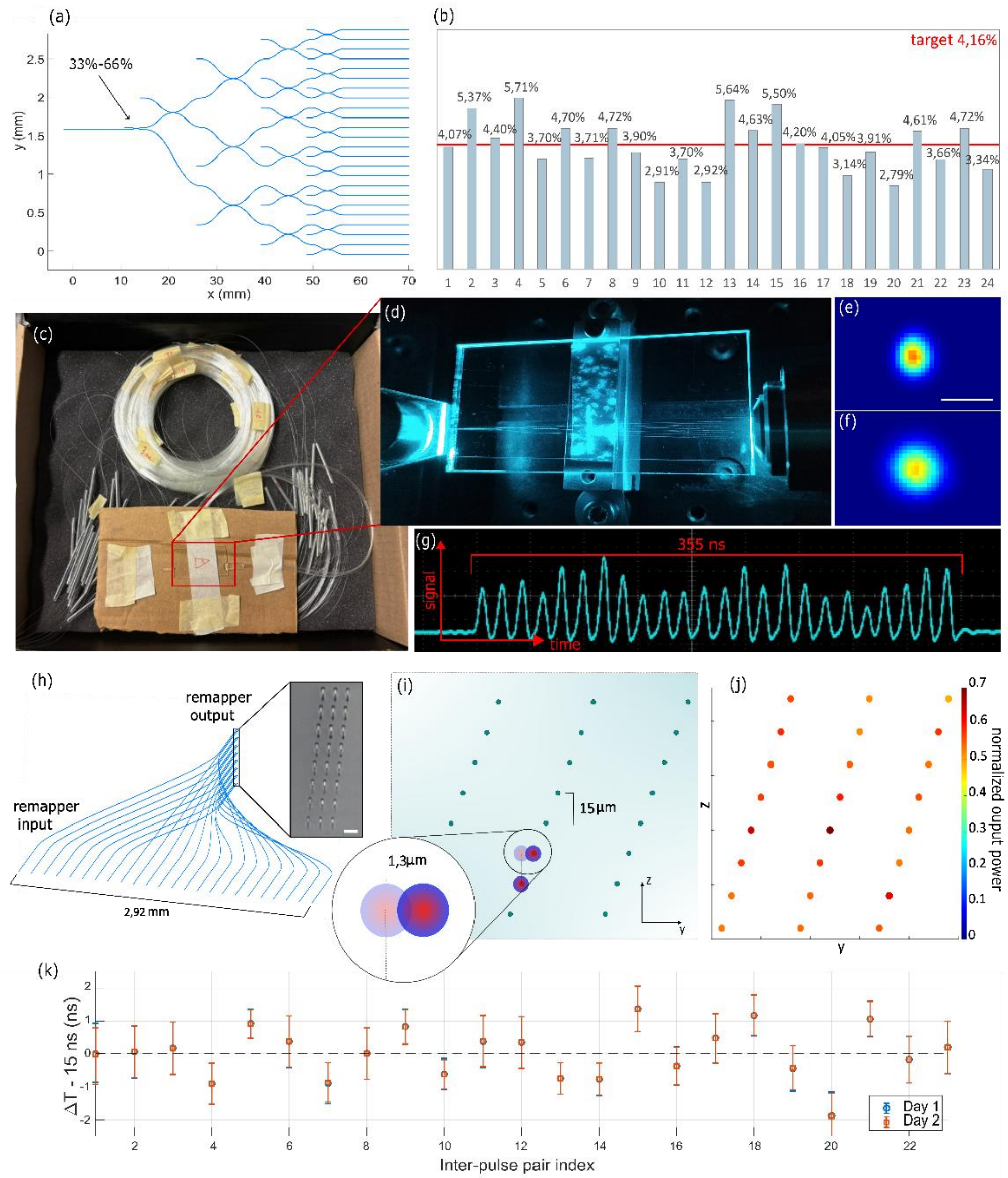

**Fig.2. Fabrication and characterization of the integrated optical components enabling spatio-temporal encoding.** (a) Design of the integrated beam-splitter network based on a cascade of evanescent couplers. The first coupler implements a 33:66 splitting ratio, followed by 50:50 stages, yielding 24 output modes. (b) Measured power distribution across the 24 outputs. The red line indicates the nominal value of 4.16%. The observed variations arise from fabrication tolerances and remain compatible with the intended application. (c) Photograph of the integrated beam splitter network coupled to the fiber-based delay lines, with a magnified view of the photonic chip when blue light is coupled in it, showing low scattering losses in the waveguides (d). (e,f) Comparison between the guided mode of the integrated waveguide (e) and the optical fiber mode (f), showing good mode matching. Scale bar: 2 µm. (g) Temporal pulse train measured at the output of the fiber delay network, showing 24 pulses separated by 15 ns over a total duration of 355 ns. (h) Layout of the 3D optical remapper. The input modes match the fiber-array outputs, while the output modes are arranged along three diagonal lines. Adjacent modes partially overlap along the y direction to ensure continuous spatial sampling and are separated by 15 µm along the orthogonal axis to prevent crosstalk. (j) Normalized output power measured at the remapper output using a single-mode fiber sequentially coupled to each waveguide. Transmission varies between 0.5% and 0.7%, mainly due to fabrication tolerances such as variations in writing depth and curvature. (k) Temporal calibration of the spatio-temporal encoder. The delay between consecutive pulses is reported relative to the nominal 15 ns spacing, showing sub-nanosecond deviations and excellent agreement between measurements acquired on different days.

Spatial encoding, in turn, is achieved using an integrated three-dimensional optical remapper, which receives the train of pulses from the fiber array and spatially reorganizes the modes to provide uniform illumination of the flowing cell. The remapper generates three parallel diagonal lines of modes lying in the *y–z* plane, all with the same inclination, which are reproduced in the center of the microchannel through a 1×1 magnification lens system. The spatial distribution of the pulses is continuous along the *y*-axis and discrete along *z*. As the cell flows through the microchannel along *z*, information is continuously and progressively acquired, enabling complete sampling of its spatial profile. Consequently, the spatio-temporally separated pulses encode two-dimensional spatial information into a single temporal waveform, allowing a single photodetector to collect the necessary information to reconstruct the complete cell image.

Figure 2 illustrates the fabrication and characterization of the different components necessary for spatio-temporal codification of the illumination signal, as described above. The integrated optical circuits (evanescent couplers and optical remapper), as well as the microchannel, were fabricated in glass by femtosecond laser micromachining [21,22]. The versatility and intrinsic 3D capabilities of this technique make it an enabling technology for the present application, where three-dimensional geometries, such as the one used for the remapper, are required. After optimizing a single waveguide for operation at a wavelength of 488 nm, achieving losses as low as 0.1 dB/cm, we fabricated the 1×24 beam splitter (Fig. 2a). Its characterization is reported in Fig. 2b, showing the power distribution at the device output compared to the target value of 4.16% of the total output power. Although some fluctuations are present due to fabrication tolerances (standard deviation is ±0.84%,), the device effectively distributes the optical power across all 24 outputs. This level of uniformity is fully compatible with our application, where maintaining signal in all modes is more critical than achieving perfect equalization. The fiber-based delay system is shown in Fig. 2c, pigtailed to the beam splitter (Fig. 2d). The optimized waveguide mode is slightly smaller than that of the fiber (2 µm vs 3 µm µm, see Fig. 2e,f), providing good mode overlap while maintaining strong confinement in the waveguide, thus preserving the spatial quality of the illumination pattern required for imaging. The pulse train measured with an oscilloscope at the fiber array output is shown in Fig. 2g, corresponding to a total duration of 355 ns. The fiber-based delay lines are then coupled to the remapper. The remapper was designed to collect at its input a fiber array collecting all the fiber delay lines (input modes are hence distributed along a single axis and spaced by 127 µm), while at the output the modes are spatially reorganized to provide uniform illumination of the sample within the microchannel along the *y*-axis (as in Fig. 2h). In this configuration, adjacent modes are partially overlapping along the *y*-axis (with a separation of 1.3 µm), while they are spaced along the *z*-axis to avoid cross-talk between waveguides (see Fig. 2i). A separation of 15 µm was found to be sufficient to prevent evanescent coupling while also ensuring good waveguide quality, since irradiation of adjacent regions during fabrication can otherwise distort previously written waveguides. This distance represents an optimal compromise between illumination compactness and uniformity. The power measured at the output of the remapper is reported in Fig. 2j, showing an average normalized transmission ($P_{out}/P_{in}$) of 0.56 ± 0.054 (mean ± SD, n = 24), corresponding to a coefficient of variation of 9.6 %. This low dispersion confirms the good uniformity and reproducibility of the remapper fabrication process. Fig. 2k shows the temporal calibration of the spatio-temporal encoder. A 1-s acquisition was performed, corresponding to ~$2\times10^6$ pulse trains. The delay between consecutive pulses was extracted and compared with the nominal 15 ns spacing. The measured mean deviation remains below 1 ns for all pulse pairs, with sub-nanosecond standard deviation. Measurements repeated on different days show nearly identical results, confirming the excellent temporal stability of the integrated delay architecture.

For the microchannel, we adopted a hybrid fabrication approach: the inlet and outlet were produced by 3D printing, while the central section, where sample illumination takes place, was fabricated in glass to ensure high optical quality of the surfaces (Fig 3.a and 3.b). The 3D-printed part was designed to perform hydrodynamic focusing of the sample, which is injected through a needle and confined at the center of the channel by a buffer solution, aligning the cells within the region intercepted by the pulse train. This hybrid configuration avoids channel constrictions and prevents clogging, which would otherwise alter the position of the hydrodynamic focus (Fig. 3b). For collecting the signal from the optical remapper and focusing it into the center of the microchannel, we used a GRIN lens (NEM-050-20-00-860-S-0.5p, GRINTECH). This commercially available lens has a diameter of 0.5 mm, a length of approximately 2.08 mm, and a numerical aperture (NA) of 0.5, making it suitable for integration with our device. It also features asymmetric object–image working distances of 0 mm and 0.2 mm in water. The lens was placed in direct contact with the remapper on one side, while on the opposite side an access hole was fabricated in front of the microchannel, allowing the lens to be positioned as close as 0.17 mm from the channel center, thus leaving a margin for experimental fine-tuning of the channel–lens distance (Fig 3.c). A 200 nm-thick gold mask, fabricated by gold deposition and femtosecond laser ablation. was deposited on the input facet of the GRIN lens to block uncoupled light from the remapper, preventing stray illumination from reaching the microchannel and affecting the spatio-temporal encoding (Fig. 3d). Figure 3e shows the assembled microchannel, while Fig. 3f presents the integration of the remapper and GRIN lens, highlighting the compact layout of the overall system.

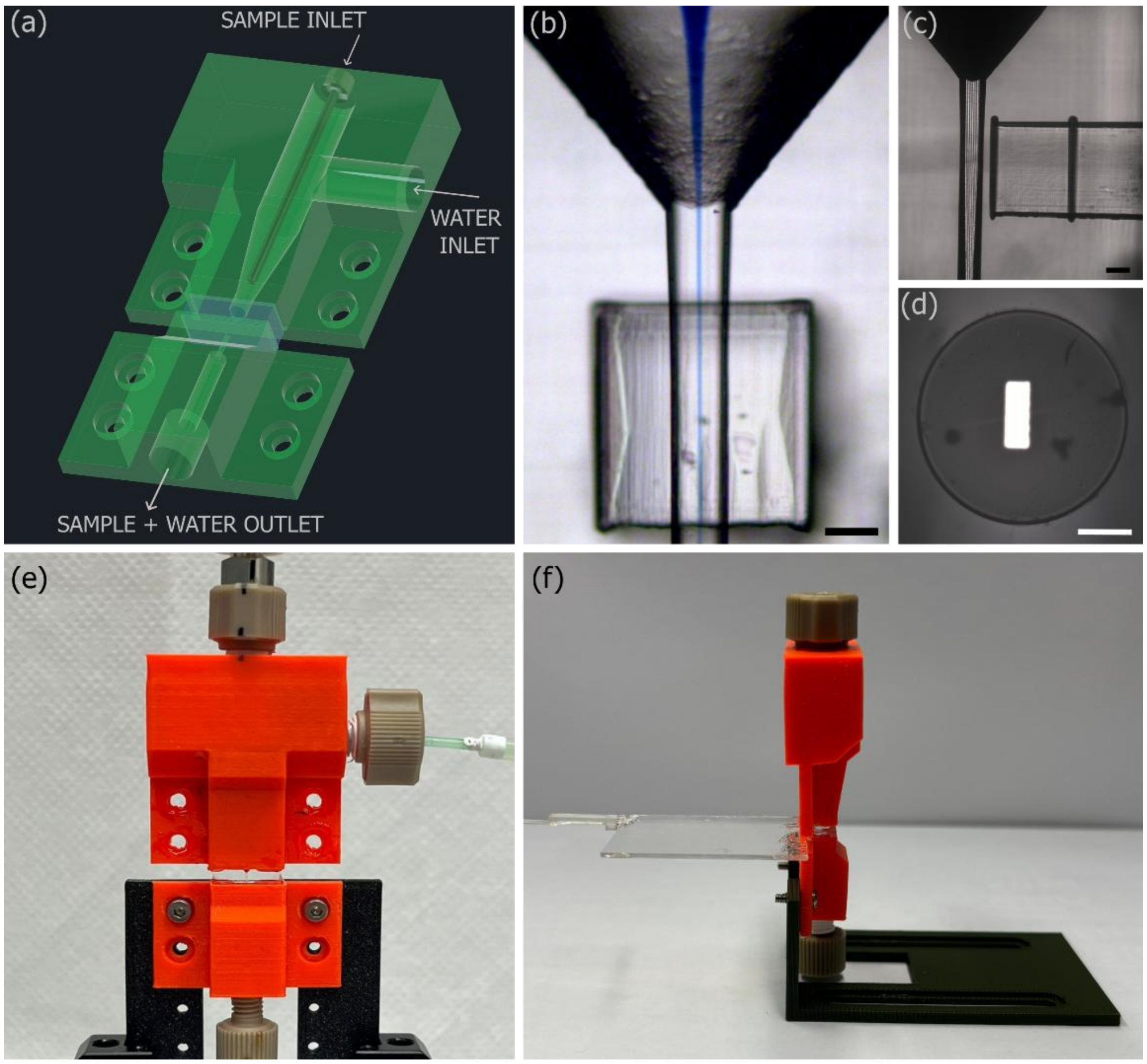

**Fig. 3. Design and integration of the optofluidic module.** (a) Three-dimensional rendering of the microfluidic device, comprising a 3D-printed component (green) and a central glass-based section, where optical interrogation takes place. The printed part incorporates two independent inlets that enable hydrodynamic focusing of the sample, confining the cells within a region smaller than the channel cross-section and aligned with the detection zone. (b) Microscope image of the glass-based section, showing effective hydrodynamic confinement. A blue dye is injected into the sample stream to visualize the focused region. The square opening orthogonal to the fluidic channel is designed to accommodate the GRIN lens, which focuses the remapper output into the center of the channel.(c) Side view of the microfluidic channel and the GRIN lens access port. (d) Image of the GRIN lens input facet, featuring a gold mask deposited to suppress background signals originating from uncoupled light at the remapper output. Scale bars: 100 µm. (e) Photograph of the assembled microfluidic device. (f) Side view of the fully assembled system, including the remapper and the GRIN lens, highlighting the compactness of the integrated architecture.

## 3. Microscope validation and discussion

The first validation of the STEPIC microscope was performed using polystyrene beads with a nominal diameter of 15 µm. During these experiments, the ratio between the inlet buffer and sample flow rates was kept constant. Specifically, the buffer (water) flow rate was set to 1 mL $min^{-1}$, while the sample flow rate was 0.03 mL $min^{-1}$, corresponding to a buffer-to-sample ratio of approximately 33:1. At this flow-rate ratio, we experimentally measured a sample confinement region of approximately 30 µm. The confinement width was quantified by imaging the thickness of a blue dye stream flowing through the microchannel. This value is consistent with theoretical estimates based on the ratio of the cross-sectional areas, assuming a square channel cross-section with a side length of 150 µm and a circular cross-section for the hydrodynamically focused sample stream. Under these assumptions, the expected confinement diameter is on the order of 30 µm, in good agreement with the experimental measurement. In addition, we measured the flow velocity of the beads under these operating conditions, obtaining a value of approximately 1 m/s. The velocity was extracted from the photodiode signal by measuring the time required for individual beads to traverse the spatial separation between adjacent illumination modes along the y-axis, equal to 105 µm. Figures 4a and 4b show typical signals acquired during bead measurements at different temporal scales. Figure 4a reports the full signal trace, where the attenuation induced by a slowly flowing bead is clearly visible, while Figure 4b shows a magnified view of the pulse train, highlighting the temporal structure of the encoded illumination. The signal is composed of a sequence of pulse trains repeating at the laser repetition rate of 2 MHz. When a bead passes through the illuminated region, the transmitted intensity of the pulses is reduced. This modulation of the transmitted signal provides the information used for the image reconstruction of the flowing object. Image reconstruction is performed using a custom MATLAB routine, whose pipeline is shown in Fig. 4c-f. In a first step (Fig. 4c), a peak-intensity variation map is generated by identifying the peaks corresponding to the 24 temporally encoded pulses and computing their relative intensity variations with respect to a reference signal. The reference is acquired immediately before sample measurement by averaging several pulse trains in the absence of beads. The resulting map highlights the intensity variations induced by the passage of a bead, appearing as a diagonal feature with non-zero values. In the example shown, the event spans 210 pulses, corresponding to a total acquisition time of 105 µs. This acquisition time corresponds to a potential throughput on the order of $10^4$ objects/s. In second and third processing steps (Fig. 4d and e), the detected pulses are reorganized by accounting for their spatial encoding and temporal ordering, enabling reconstruction of the two-dimensional image of the bead as it flows through the microchannel. Additional pixel interpolation, and filtering are applied to obtain the final reconstructed image, shown in Fig. 4f. Figures of different beads shown in grey colormap are reported in Fig. 4g. The average bead dimension retrieved from reconstructed images is about 16 μm, consistent with the nominal bead size and the system resolution.

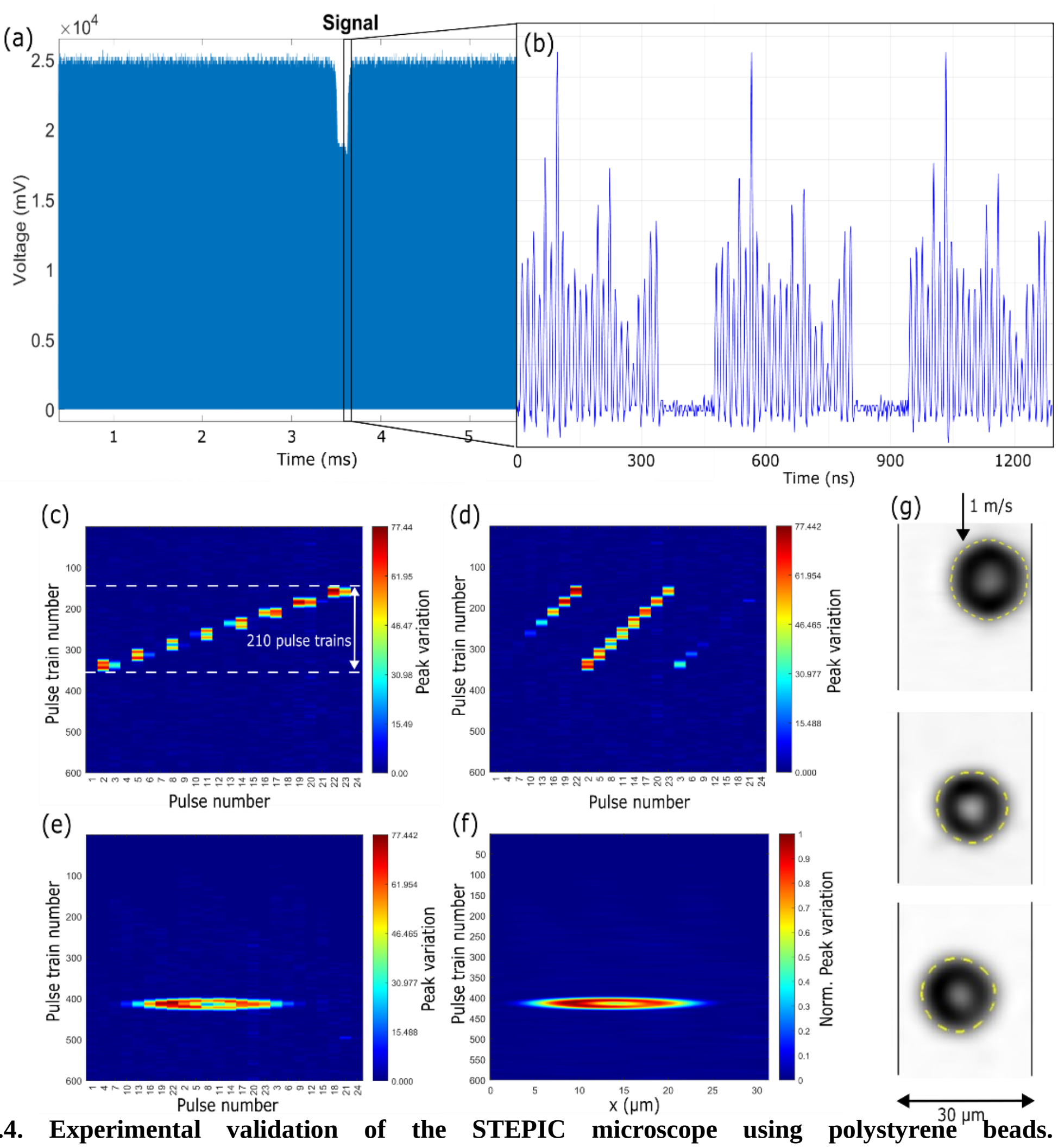

**Fig.4. Experimental validation of the STEPIC microscope using polystyrene beads.** (a) Representative pulse train acquired by the single-pixel photodiode, showing the intensity modulation induced by the passage of a bead through the detection region. (b) Detail of three consecutive pulse trains, highlighting the temporal encoding of the spatial information. (c–f) Illustration of the image reconstruction procedure. (c) Peak-intensity variation matrix, where each column corresponds to one of the 24 spatial modes and each row corresponds to a temporally successive pulse train. Each matrix element represents the relative intensity variation with respect to a reference signal acquired in the absence of flowing sample. The diagonal feature indicates the transit of a bead through the detection region, spanning 210 pulse trains and corresponding to an acquisition time of 105 μs, consistent with a flow velocity of approximately 1 m s$^{-1}$. (d) Matrix after spatial decoding, which accounts for the physical position of each mode in the sample plane. (e) Matrix after temporal decoding, exploiting the known inter-pulse delays to recover the correct temporal ordering. (f) Final reconstructed image after noise filtering and pixel interpolation. (g) Examples of reconstructed bead images displayed in grayscale. Black side edges limit the field of view.

The bead experiments allow discussing the key performance metrics of the STEPIC microscope, including spatial resolution, temporal sampling, and achievable throughput. The spatial resolution of the present system is primarily determined by the size of the illumination modes projected onto the sample plane. In the current implementation, the mode-field diameter at the microchannel corresponds to approximately 2 µm, which sets the fundamental optical resolution limit. The effective spatial sampling is defined by the spacing between adjacent illumination modes along the transverse direction (1.3 µm) and by the temporal sampling along the flow direction, which depends on the cell velocity and the pulse repetition rate. Under the operating conditions used in this work, these parameters allow reliable reconstruction of micron-scale morphological features of flowing objects. In the present architecture, the trade-off between spatial resolution and throughput is determined by the number of encoded illumination points and by the temporal spacing between pulses. Increasing the number of spatial sampling points improves image resolution but proportionally increases the acquisition time. However, this trade-off is governed by the encoding architecture rather than by detector limitations, allowing straightforward scalability of the system. In particular, higher repetition-rate laser sources or shorter optical pulses would directly increase the spatio-temporal sampling density, enabling improvements in both temporal resolution and throughput without modifying the underlying on-chip architecture.

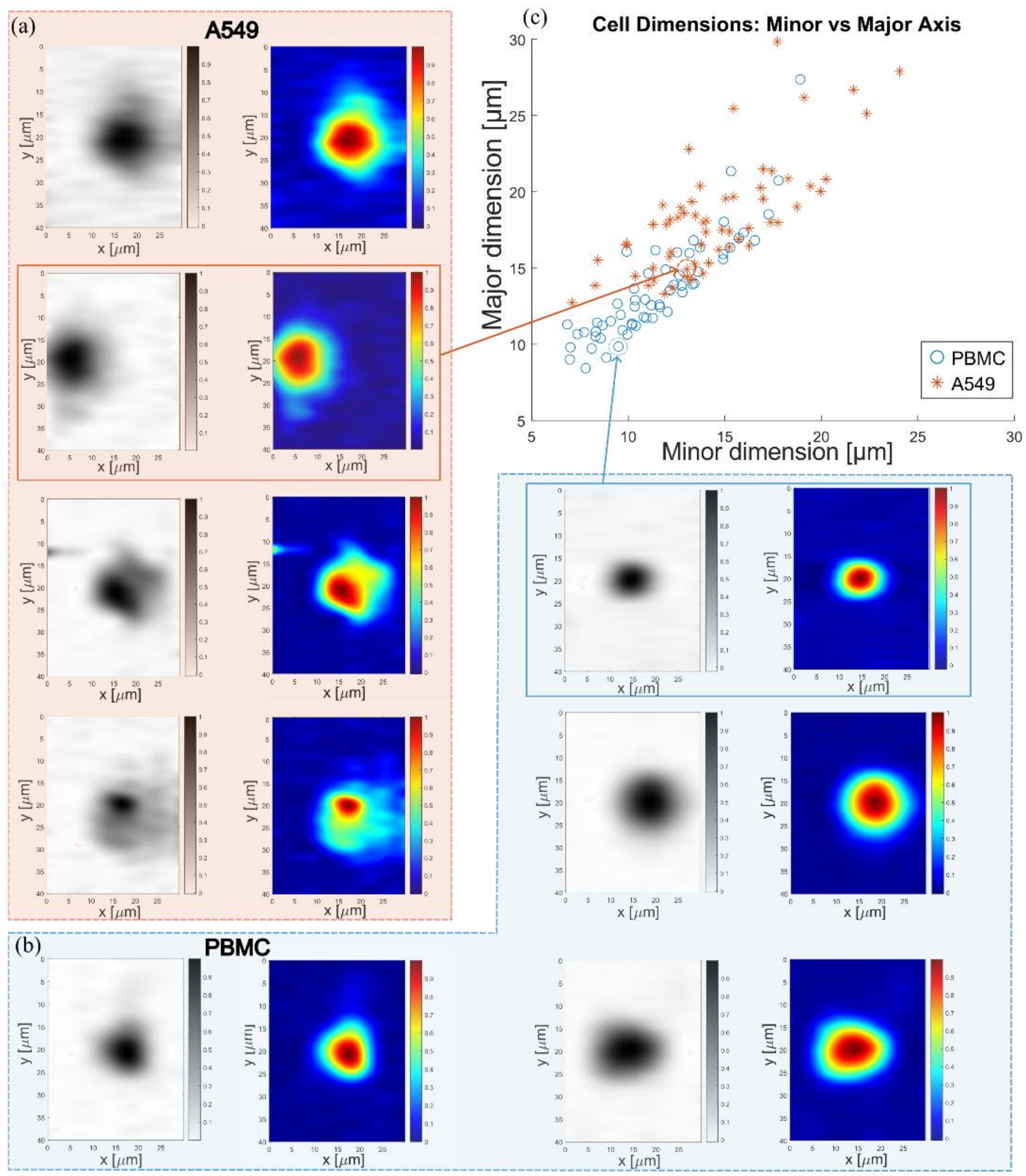

**Fig. 5. Device validation using distinct cellular populations.** Representative images of the A549 cell line (a) and peripheral blood mononuclear cells (PBMCs) isolated from healthy donors (b), shown both in grayscale and using the jet colormap. Grayscale images provide a faithful representation of the measured signal, while the jet colormap enhances the visualization of morphological features and local intensity gradients, improving qualitative interpretation. Panel (c) reports the size distribution of the two populations, showing for each cell the lengths of the minor and major axes.

With the final aim of optimizing our integrated microscope for automated CTC identification, we tested tumor and immune cell samples mimicking patient blood samples. Following the initial bead acquisition, we processed different cell populations, including the lung epithelial cancer cell line A549 and peripheral blood mononuclear cells (PBMCs) from healthy donors. From the reconstructed images, we extracted two orthogonal cell diameters along the vertical and horizontal directions of the reconstructed image. The larger and smaller of these values were used as estimates of the major and minor cell diameters, respectively. The results are summarized in Fig. 5, together with representative reconstructed images for each population. These simple geometric descriptors provide a first quantitative characterization of the reconstructed cell images and allow comparing the typical size ranges of different cell populations. This analysis illustrates the capability of the system to retrieve morphological information from flowing biological specimens and highlights its potential for future applications in high-throughput cellular analysis.

## 4. Conclusions

In this work, we have demonstrated the first fully integrated microscope enabling high-temporal-resolution imaging based on spatio-temporal encoding in photonic integrated circuits. By exploiting the three-dimensional capabilities of femtosecond laser micromachining, we achieved an unprecedented level of optofluidic and photonic integration, resulting in a compact and mechanically stable platform. This high degree of integration leads to two major practical advantages. First, the system enables straightforward operation, requiring only standard syringe pumps for sample injection. Second, the monolithic integration and permanent alignment of the optical components provide exceptional system stability, reducing sensitivity to day-to-day alignment variations that typically affect free-space implementations and strongly enhancing measurement reproducibility Such stability is particularly advantageous for machine-learning based analysis pipelines, which rely on consistent input data distributions for reliable training and classification of cellular populations. The performance demonstrated in this work establishes a new reference point for fully integrated ultrafast imaging systems. In the current implementation, a single image is acquired in approximately 100 µs, corresponding to an effective throughput of $10^4$ objects per second. This acquisition time is not an intrinsic limitation of the STEPIC architecture, but rather reflects the characteristics of the employed laser source, which delivered 10 ns pulses at a repetition rate of 2 MHz. The platform is fully compatible with shorter optical pulses and higher repetition rates, enabling straightforward scaling toward substantially higher temporal resolution and throughput. Similarly, the present spatial resolution, on the order of 2 µm, is primarily determined by the mode-field diameter of the integrated waveguides and by the optical magnification of the focusing optics. Both parameters can be readily engineered. For instance, employing a larger number of spatial modes (e.g., 48) combined with a GRIN lens providing 0.5× magnification would allow imaging of specimens with diameters of ~30 µm at ~1 µm spatial resolution, while preserving acquisition times of approximately 200 µs, or significantly shorter when combined with reduced pulse durations. Overall, STEPIC microscopy establishes a scalable and robust platform for ultrafast, high-throughput imaging, bridging the gap between laboratory-scale ultrafast microscopes and practical deployable instruments. By uniting microfluidics and integrated photonics in a single monolithic architecture, this approach opens the route toward compact imaging flow cytometers for real-time, large-scale biological analysis, with potential applications ranging from studies of cellular heterogeneity to the detection of rare circulating tumor cells (CTCs).

## 5. Methods

### A. Fabrication of optical and fluidic components

Integrated photonic structures were fabricated in borosilicate glass substrates (EAGLE XG, Corning) using femtosecond laser micromachining (FLM), which enables the precise three-dimensional realization of integrated components. This capability is particularly suited to the present work, as it allows the fabrication of structures required to generate the tailored and 3D illumination patterns used in the STEPIC technique. Fabrication was performed using two commercial femtosecond laser systems (Carbide and Pharos, Light Conversion) due to different system availability during the device fabrication process. Both systems are based on DPSS Yb-doped architectures and operating at a central wavelength of 1030 ± 10 nm, we have used 180 fs pulses and 1 MHz repetition rate for all fabrication processes. Waveguides were irradiated about 400 µm below the surface using an Olympus LCPLN50XIR objective (50×, 0.65NA, with a correction ring for compensating spherical aberrations up to 1.2 mm), with writing parameters of 190 mW and 8 scans for the Carbide laser at 230 mW laser power, while 12 scans were used with the Pharos laser. High-precision sample translation was provided by an Aerotech FIBER-Glide 3D positioning system, offering 2 nm resolution and ensuring smooth and accurate motion during writing. After irradiation, the chip facets were polished and subsequently annealed in oven up to 750 °C to reduce stress-induced birefringence, using a receipt previously optimized for propagation losses minimization [23,24]. With this process we have obtained single modes waveguides at 488nm, with 0.1 dB/cm of propagation losses.
We subsequently optimized the layout of the directional couplers to achieve the desired splitting ratios. To this end, we fabricated a series of test couplers with increasing waveguide separation, ranging from 3 to 5.1 µm, while keeping the interaction length fixed at 0 µm. The radius of curvature was set to 45 mm in order to minimize bending losses. By fitting the experimentally measured coupling data, we determined optimal arm separations of 3.18 µm and 3.45 µm to obtain the targeted 33:66 and 50:50 splitting ratios, respectively.
Microfluidic channels were fabricated using the FLICE (Femtosecond Laser Irradiation followed by Chemical Etching) technique on fused silica substrates, enabling the formation of hollow three-dimensional structures within the glass. For this process, the second harmonic at 515 nm was used for sample irradiation, with a laser power of 550 mW, focused with a 50× 0.6 NA microscope objective. Scan speed was kept equal to 1 mm/s. Wet etching was then performed in a 20% HF solution under ultrasonic agitation. To generate the metallic mask over the grin lens input facet, a 200 nm gold layer was deposited and patterned by laser ablation (220 mW, 2 scans) using a 10×, 0.25 NA microscope objective. Device assembling has been achieved using 6-axis computer controlled high-precision micromanipulators (Hexapods, PI) and once the coupling was optimized the components have been glued with a UV curable glue (DELO Photobond).

### B. Optical and fluidic device characterization

Optical characterization was performed in a butt-coupling configuration, where the device was aligned between a single-mode input fiber and a collection objective. A continuous-wave laser at 488 nm (COHERENT® OBIS 488LS) was coupled into single-mode fibers mounted on a V-groove holder fixed to a 3-axis micromanipulator (Thorlabs NanoMax MAX313D) for fine input alignment. The chip was positioned on a 4-axis stage (Melles Griot 17AMT001/D) enabling accurate translational and rotational alignment with respect to the fiber. Light emerging from the waveguides was collected using a Leica 20×, NA 0.30 objective and directed either to a power meter (Ophir Nova II) or to a CCD camera (Thorlabs DCU224M). The microfluidic functionality of the chip was evaluated using two syringe pumps (KDS410, from KDScientific, and Pump Elite 11 from Crisel Instruments) connected to the fluidic tubes used for buffer and samples to independently control the flow rates of each channel. The device was observed under an optical microscope (LEICA DMI3000M) while injecting colored aqueous solutions, allowing the visualization of the fluid streams inside the microchannel. By adjusting the flow rates of the two pumps, it was possible to characterize the hydrodynamic behavior of the system and identify the operational conditions required to achieve a stable and confined flow regime within the central region of the channel. For the final assembly of all device components, it was necessary to interface the different elements using ferruled fibers and fiber arrays, and to accurately align the optical and microfluidic sections of the system.

**C. Sample preparation**

The beads used were fluorescent beads from Spherotech (FP-15056-2). The sample was prepared diluting 50 µl of polystyrene beads in 2 ml of water solution. Tumor cells were obtained from the A549 lung adenocarcinoma cultured in vitro in RPMI medium supplemented with 10% fetal bovine serum (FBS). Cells were detached by trypsinization, counted using the trypan blue exclusion method, washed in PBS 1x with calcium and magnesium (Sigma), and resuspended at a final concentration of $2\times10^6$ cells/mL in PBS 1x. Peripheral blood mononuclear cells (PBMCs) were isolated from 10 mL of peripheral blood obtained from healthy donors using SepMate™ PBMC Isolation Tubes according to the manufacturer's instructions. PBMCs were then counted and resuspended at a final concentration of $2\times10^6$ cells/mL in PBS 1x.

**D. Data acquisition and processing**

Pulse train for image acquisition was acquired with a fast photodiode (PDA10A2, Thorlabs) and processed with a high speed dizitizer (M4i.2210-x8 PCIe Digitizer, Spectrum Instrumentation). Custom Labview and Matlab softwares are used for signal acquisition and subsequent processing, respectively.

**Acknowledgment** Paiè P. thanks European Union – Next Generation EU - "PNRR - M4C2, investment 1.1 - "Fondo PRIN 2022" – High Throughput microscope on a chip for liquid biopsy assisted by artificial intelligence, HIMALAYA – id 20228MHWPZ – CUP D53D23001090001 and EXPLORER a MSCA-SE European project (id 101235654). The authors acknowledge Serendipity Lab for fostering the international collaboration and JST ASPIRE for supporting the collaborative activities.

**Disclosures.** The authors declare no conflicts of interest.

**Data availability.** Data underlying the results presented in this paper are not publicly available at this time but may be obtained from the authors upon reasonable request.